\documentclass{svproc}
\usepackage{url}

\usepackage{xcolor}
\usepackage{graphicx}
\usepackage{lmodern}
\usepackage{tcolorbox}
\usepackage{caption}
\usepackage{subcaption}
\usepackage{amsmath, amssymb}

\begin{document}
\mainmatter              
\title{Extending the Bayesian Framework from Information to Action}
\titlerunning{Extended Bayesian methods}  
%
\author{
Vasileios Basios\inst{1}, Yukio-Pegio Gunji\inst{2} \&
Pier-Francesco Moretti\inst{3} 
}
\authorrunning{Basios, Gunji, Moretti (2023)} 
%
\tocauthor{
Vasileios Basios, Yukio-Pegio Gunji, Pier-Francesco Moretti, 
}
\institute{Service de Physique des Systèmes Complexes et Mécanique Statistique 
and Interdisciplinary Center for Nonlinear Phenomena and Complex Systems 
C.P.231 CeNoLi-ULB, Université Libre de Bruxelles (ULB), Brussels, Belgium.\\
\email{vasileios.basios@ulb.be},
\and
Department of Intermedia Arts and Science, School of Fundamental Science and 
Technology, Waseda University, Tokyo, Japan.\\
\email{gunji@uwaseda.jp}
\and
CNR, National Research Council, P.le A. Moro 7, Rome, Italy.\\
\email{pierfrancesco.moretti@cnr.it}\\
}

\maketitle              

\begin{abstract}
In this review, we examine an extended Bayesian inference method and
its relation to biological information processing. We discuss the idea of
combining two modes of Bayesian inference. The first is the standard Bayesian
inference which contracts probability space. The second is its inverse, which
extends and enriches the probability space of latent and observable variables.
Their combination has been observed that, greatly, facilitates discovery. 
Moreover, this dual search during the updating process elucidates a crucial 
difference between biological and artificial information processing. The latter 
is restricted due to nonlinearities, while the former utilizes it. 
This duality is ubiquitous in biological information process dynamics 
(`flee-or-fight', `explore-or-exploit' etc.) as is the role of 
fractality and chaos in its underlying nonequilibrium, nonlinear dynamics. We 
also propose a new experimental set up that stems from 
testing these ideas.


\end{abstract}

\section{Introduction}
\label{intro}
\begin{quote}
\emph{“Explore different areas.The statement that one cannot be 
both deep and broad is a myth. Actually, the importance of being a polymath is 
that it allows one to make remote associations, and thus to understand the 
deeper essence of things. Understanding is nothing more than elucidating 
associations.”} -- A. Fokas \cite{7secrets}.
\end{quote}

In its wide range and seminal work of Professor Athanasios Fokas one finds 
important contributions in an interdisciplinary research 
fashion on the interface between applied and pure mathematics. 
Among other important contributions, Professor Fokas worked and taught a lot 
about the challenge that inverse problems pose: how to hypothesise and 
determine the most plausible set of causal interconnections and identify the 
physical and/or statistical laws that govern the acquired data. 

We too, find inspiration and try our best to follow his tall 
order of `exploring different areas'. And in that spirit we  
approach this interdisciplinary area of biological information 
processing. In particular, we trace the development of the idea that chaos, 
affording indeterminacy and criticality, is the `condicio sine qua non' 
for biological information processing and we highlight the 
importance of fractal basin boundaries for such an interpretation. 

Moreover, we shall see how this entails a nonlinear closed feedback loop with  
the forward and inverse inference problems interlaced. 
The 'forward problem', in this case is that of categorization, i.e to propose 
models that calculate the results from, and reactions to, their causes. In our 
case it is the framework of the classical Bayesian theorem. The `inverse 
problem' here deals data acquisition, i.e to calculate causes from results. And 
in our case connects with the converse of the Bayesian theorem, a fertile but 
less travelled road, which expands its original, classic, framework
\cite{BIB_Frontiers2022,BasiosGunji2017,GunjiBasios2017,GunjiSwarm2018},
\cite{Shinohara2020,BasiosGunji2021}.

The paper is organized as follows: 
in Section \ref{sec:bio_vs_art}  we review  
aspects that differentiate biological from  
artificial information processing, and dynamics, focusing on the role of 
chaos and fractals. 
In Section \ref{sec:bayes} we discuss the connection of Bayesian inference to 
the free energy principle and present an outline of our proposed extension.
With Section \ref{sec:conclusion} we conclude with a short discussion of 
forthcoming research plans and their outlook.

\section{Biological vs Artificial Information Processing}
\label{sec:bio_vs_art}
Life is abundant with information processing, this is commonplace, in our era 
though information flow is not bound to life's biological processes. We are 
surrounded by artificial technological contraptions serving, processing and 
acquiring information. Some of them are bio-inspired some are based in Turing's 
and Shannon's prototypical mechanistic theories of information. Evidently 
natural systems differ in structure and function from human-made computers 
as much as natural patterns differ from human-made ones. As the father of 
Fractal Geometry  Beno\^it Mandelbrot famously put it \emph{``Clouds are not 
spheres, mountains are not cones, coastlines are not circles, and bark is not 
smooth, nor does lightning travel in a straight line"}, and biological 
information processing is not machine-information processing; and fractals do 
appear here, too.

\begin{figure}[h]
\centering
\includegraphics[width=7cm]{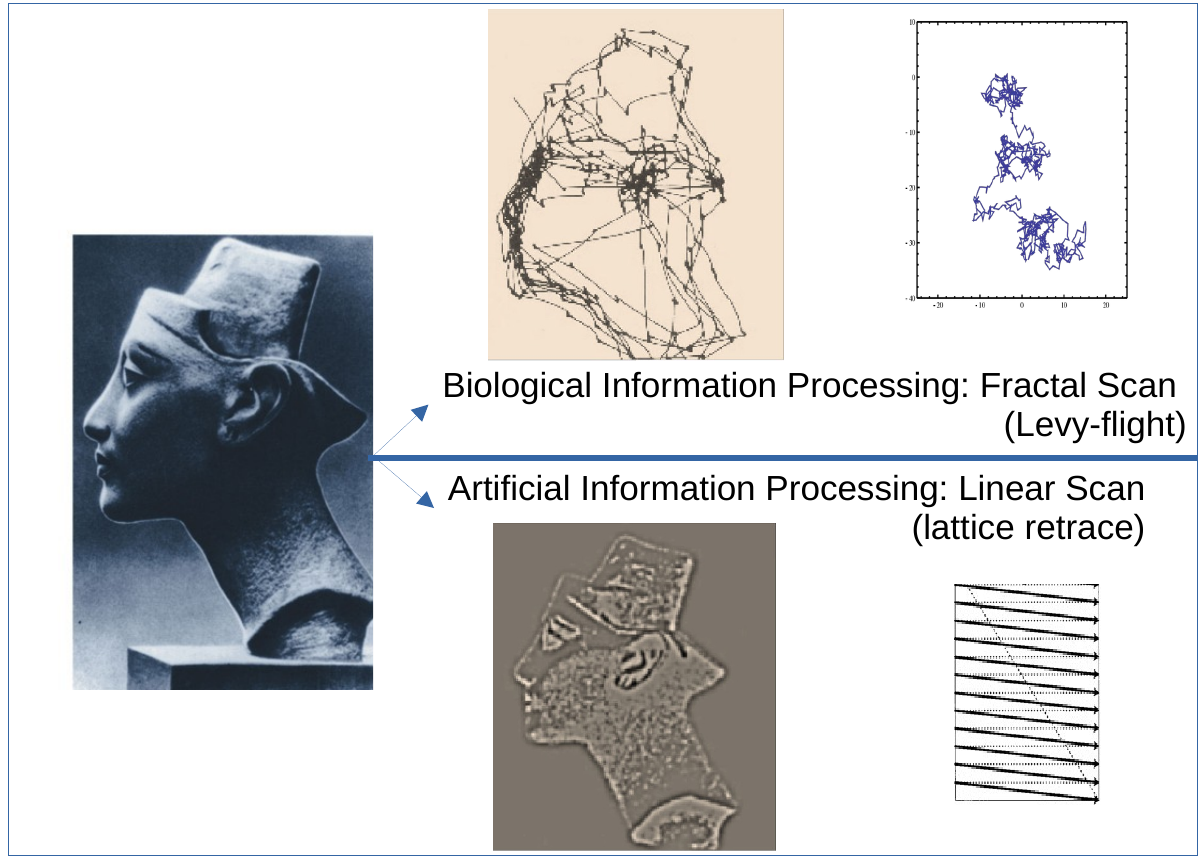}
\caption{`clouds are not spheres, mountains are not cones' seeing is not 
scanning}
\label{fig:nefertite}
\end{figure}

From the early times of information machinery it became evident that 
acquisition and storage of information in biological systems happens in a 
radically different way. Take for example the first historical study of how 
human visual system scans an image. In humans it happens in a fractal itinerary, 
what it came to be known as a Levy flight \cite{JSN_Tsuda1985}. 
In a scanner it happens via a serial-sweep on a 
lattice, as Fig.\ref{fig:nefertite}.
Moreover, biological systems process information at a multitude of levels.
Or, as a pioneer of the subject, John S. Nicolis,
put it \cite{JSN_GN_VB_2015} \emph{``The smallest biological information 
processor is the enzyme; the 
biggest is the (human) brain. They are separated by nine orders of magnitude. 
Yet their complexity is comparable"}. Hence, fractality is obviously the 
most compatible property of choice for such cases of distributed 
probabilistic computations as the ones encountered from proteins, to organs and 
organisms, even to groups of organisms.

\noindent{\bf The Role of Chaos, Fractals \& Complexity}:
It has been established,since the early days of Chaos theory that a reliable 
information processor must allow for chaos \cite{JSN_Tsuda_1999}. In particular 
it must 
afford the regimes known as 'edge of chaos' or `self organized 
criticality' that are ubiquitous in systems with coexistent negative and 
positive nonlinear feedback circuits. Chaos Theory has shed new light in 
phenomena associated with biological information processors, see for example 
\cite{Poil2012,Chialvo2010}.
Such type of chaotic dynamics allows also for adaptivity, flexibility and 
resilience during information processing.
Moreover, since biological information is contextual, meaningful, has depth 
of memory and historicity. Other related types of chaotic dynamics, such  
stochastic resonance, intermittency and chaotic itinerancy were also found to 
play key roles \cite{JSN_Tsuda_1999,Tsuda2015}.

The role of chaos in biological systems is exemplified: (i) at the 
macroscopic level in the analysis of chaos-order transitions in the brain and 
other organs (ii) at the microscopic 
level, modeling of neurons as systems of nonlinear differential equations, 
even revealing new dynamics (blue sky catastrophe and spike-trains), and 
(iii) at the mesoscopic level in groups of neuron communities where 
chimera states, non-local synchronization and modular collective dynamics were 
identified. So, it comes with no surprise that even the new trends of 
bio-inspired Information processing paradigms (e.g. artificial neural 
networks) discover the constructive role of chaos.

\noindent{\bf The importance of being Fractal \& Chaotic:}
The essence of biological information processing can be expressed as the 
emergence of nonlinear feedback loops with two branches: The branch that 
provides stability and facilitates data storage, i.e. 
data-categorization. This is modelled via attractors in the phase space with 
dynamics characterized by a negative Lyapunov exponents' sum, $\Lambda <0$. 
While the other branch provides instability and facilitates data acquisition 
actions, i.e. observation. This in turn is modeled via chaotic exploration of 
the phase 
space, with dynamics characterized by a positive Lyapunov exponents' sum, 
$\Lambda >0$, \cite{BasiosGunji2017,JSN_Tsuda_1999}. 

We must take note that the physical and biological parameters, 
here, are not always constant in time and a more complete treatment, albeit much 
more complicated and demanding would have to account for `chaotic itinerancy'.
Chaotic itinerancy \cite{JSN_Tsuda_1999,Tsuda2015} is a quite generic 
mechanism for high dimensional systems with coexisting fast-slow dynamical 
subsystems. It captures the complexity, plasticity and flexibility of 
biological systems, particularly neural dynamics, and since it is contingent 
upon history and parameter switching allows transitions to be stochastic 
\cite{Poil2012,NicNic2012}.

This emergent feedback loop is reminiscent of the `fight or flee', `exploit or 
explore' phenomena in biological systems at large. Only here the 
negative-feedback, contracting, branch (fight/exploit) has to do with 
comprehension via pre-existing categories, while the positive-feedback, 
expanding, branch has to  do with seeking the knowledge of new data. 
Moreover a well functioning loop has to be well tuned, poised on the border 
of chaos and order.
The following scheme summarizes this ubiquitous closed feedback loop pair:
\begin{tcolorbox}[width=12cm]
{\bf Categorization:} exploiting of phase space. \\
Stability, $\Lambda$  $<0$ $\longmapsto$
Data Storage/Memory: Inhibitory modes\\
{\bf Observation:} exploring in phase space. \\
Chaos, $\Lambda >0$ $\longmapsto$
Data Acquisition/Input-Output: Excitatory modes
\end{tcolorbox}
\noindent or as J.S. Nicolis and I. Tsuda put it \emph{``To observe you need 
a priori categories, but to form categories you need observations"} 
\cite{JSN_Tsuda_1999}.

\begin{figure}
     \centering
     \begin{subfigure}[b]{0.33\textwidth}
         \centering
         \includegraphics[width=\textwidth]{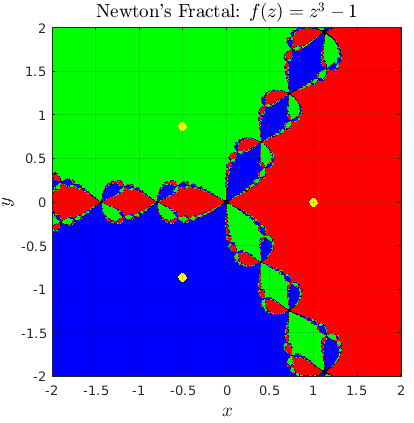}
         \caption{An example of three coexisting attractors' basins 
with their fractal basin boundaries. Each cubic-root's of unity (yellow dots), 
basin is in different color.}
         \label{fig:basinNewton}
     \end{subfigure}
     \hfill
     \begin{subfigure}[b]{0.15\textwidth}
         \centering
         \includegraphics[width=\textwidth]{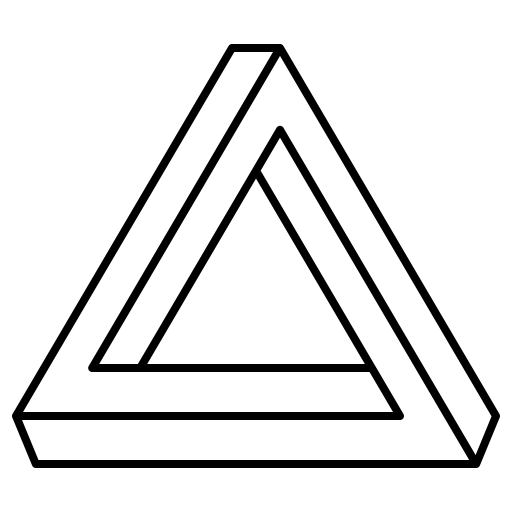}
         \caption{ } 
         \label{fig:PenroseTriangle}
     \end{subfigure}
     \hfill
     \begin{subfigure}[b]{0.5\textwidth}
         \centering
         \includegraphics[width=\textwidth]{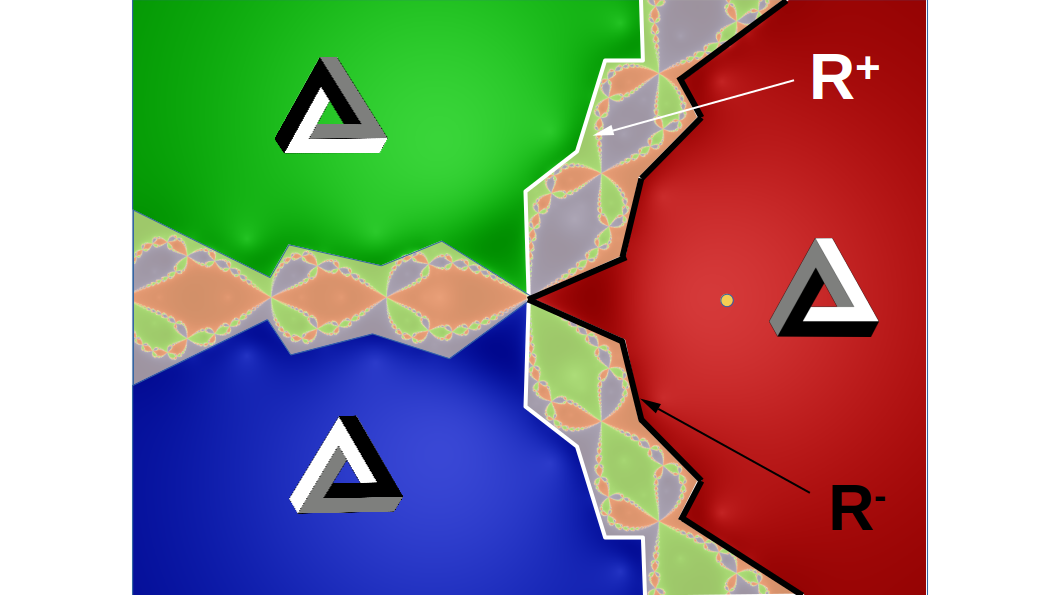}
         \caption{The three coexisting attractors with coarse grained 
(rough-set) boundaries modelling the three switching perceptions of Penrose's 
Triangle of Fig.1(b). The fractal basin boundary is within the highlighted 
region. The $R^{\pm}$ lines delineate roughly the basin around the root 
$z_1=1$. (yellow dot). } 
         \label{fig:Trillusion} 
     \end{subfigure}
     \hfill
        \caption{Coexisting attractors with fractal basin boundaries in 
biological information processing provide models for multistable 
perception and semantic/categorical polyvalence.}
        \label{fig:basin_boundaries}
\end{figure}

\noindent {\bf Fractal Basin Boundaries:}
Another key aspect of chaotic dynamics is the coexistence of attractors. And it 
is well known that these coexisting attractors are, most often than not, 
separated by fractal basin boundaries \cite{BasiosGunji2017,NicNic2012}. One 
of the most well known systems with 
this property is the celebrated `Newtons Fractal' a Julia set associated to 
Newton's iteration method for finding roots of polynomials $f(z):  z_{n+1} 
:= z_n -\frac{f(z_n)}{f'(z_n)}, z \in \mathbb{C}$. For degree 3, $f(z)=z^3-1$ 
the attracting roots of unity and their fractal basin boundaries are shown in 
Fig.\ref{fig:basin_boundaries}. Models with fractal basin boundaries have been 
proven to capture well the affinity of concepts in their apprehension and 
categorization and the intermittent successions among them. This kind of 
multistability and its stochastic switching is exemplified in the study of 
optical illusions. Bistable transitory dynamics, such as the ones found in the 
perception of the Necker Cube, or Wittgenstein's favourite linguistic paradigm 
of the Rabbit/Duck picture, were of the first to highlight the importance of 
inhibitory and excitatory connections in neural correlates during perception.

Furthermore, fractal basin boundaries of coexisitng (strange or not) attractors 
provide also for the probabilistic, stochastic, aspect inherent in biological 
processors. As it is well known a chaotic system under coarse-graining cannot 
be distinguished from a stochastic one \cite{NicNic2012}. This is due to two 
facts: 
Firstly, the practical impossibility to fully determine initial conditions, or 
any point of the phase-space for that matter, with infinite accuracy. Secondly, 
the inherent sensitive dependence on initial conditions of chaotic systems 
which renders unpredictable the course of their evolution beyond their Lyapunov 
time ($\tau \approx 1/\Lambda)$. Hence, for any point on the fractal basin 
boundary, which is necessarily determined with finite accuracy, we can only 
assign a probability, weighted by the boundary's fractal measure, of 
arriving at a neighbouring attractor of the coexisting ones.

It is customary to partition the phase-space of a dynamical system by the 
preimages, periodic points and/or other critical points, or by a simple lattice 
that is refined iteratively, as in the cases of fractal-dimension determination 
by box-counting. For example, in Fig.\ref{fig:basin_boundaries} the highlighted 
region contains the fractal boundary. We can then use a rough set or a coarse- 
graining scheme, e.g. the partition determined by the 
curves $R^+$ and $R^-$ marked with the white and black lines in 
Fig.\ref{fig:basin_boundaries} that enclose the fractal boundary of the basin 
of attraction of the first cubic-root of unity, in this case, ($z_1 = 1$). 
Every point inside $R^-$ will end up in $z_1$, every point outside $R^+$ will 
\emph{not} end up in $z_1$ while any point in the (highlighted region) could 
end up either end up in $z_1$ with a given probability, determined by the 
underlying fractal measure, or end up in \emph{either} $z_2$ or $z_3$, the 
other two cubic-roots of unity. Similar argument holds for the other two 
attractors around $z_1$ and $z_2$, with respective partitions. 

Fig.\ref{fig:Trillusion} illustrates the above mechanism that gives rise to a 
tri-stable visual perception of the famous Penrose Triangle 
shown in Fig.\ref{fig:PenroseTriangle}. When the visual cortex is impinged upon 
with such ambiguous stimuli, it is impossible to categorize it in a single 
perspective. So, all three pre-existing possible and competing categories of 
perspective are excited. The inhibitory part of the circuit drives the 
system towards one of these three, but in a probabilistic fashion. Because the 
uncertain data on the fractal basin provides the fluctuations for stochastic 
transitions from one category/fixed-point to the other. Note that, such a 
loop puts data (stimuli) and representations (attractor-basins) on a shared 
basis.

These typical far-from-equilibrium processes are directly related to the 
ever present dissipative structures' dynamics in biological systems 
\cite{NicNic2012}.
It results in a measurable effect with well determined 
transition probabilities, for a detailed model of tri-stability see 
\cite{IanStewart2019}. 
Indeed ambiguous figures are found as guiding paradigmatic cases in every other 
textbook in cognitive sciences and neuroscience as a `gateway to perception'. 
They are markedly elucidating cases of the dynamics of cognition since 
here the perception changes but not the signal. Which echoes another's pioneer 
take on the subject, Walter Freeman's, , who stated that: 
\emph{``perception depends dominantly on expectation and marginally on sensory 
input''}, see his contribution in \cite{JSN_GN_VB_2015}.

Moreover, it has been established 
\cite{BIB_Frontiers2022,GunjiBasios2017,GunjiBasios2016}
that such transition between 
interrelated categories, as exemplified here, when treated within an 
extended Bayesian framework give rise to a logic with clear non-Boolean 
characteristics in accordance with the theory of Quantum Cognition 
\cite{QC2010,QC2012,Aerts}.

\section{Extended Bayesian Inference: two modes in one loop}
\label{sec:bayes}
If biological, or even bio-inspired, information processing   
did not have this emergent nonlinear loop of inhibitory/excitatory dynamics as 
its defining characteristic, then classical Bayesian 
inference would suffice to describe categorization, the stable part, as it 
is implemented via computers' mechanical information processing.
Bayesian inference is formally equivalent with a variational free-energy 
principle minimization problem or a `least action variational problem'
\cite{BIB_Frontiers2022,Friston2012,Friston2010}. 
It is a classical optimization problem encountered in statistical 
mechanics and thermodynamics among other disciplines. It is often pictured as 
climbing up \emph{one} mountain top (or equivalently descending in \emph{one} 
valley basin depending on the choice of sign). 
\\
\noindent {\bf The Free Energy Principle in cognition \& action:}
One way that Friston and co-workers \cite{Friston2012,Friston2010} express this 
fact is by using 
the following succinct formula for the free energy functional, $F$: 
\begin{equation}
 F(q(s),p(\mu);\eta) = Energy-Entropy = -\langle \ln p(s;\eta)\rangle_q + 
\langle \ln 
q(\mu;\eta) \rangle_q 
\end{equation}
\noindent where $p(s;\eta)$ and $q(\mu;\eta)$ are 
probabilistic representations (i.e variational densities) of sensory inputs, 
$s$, from the environment, and the system's internal representations, $\mu$,
both weighted with respect to the system's external states, $\eta$ and 
conditioned over $q$. Other, alternative, expressions are derived and presented 
in \cite{Friston2012,Friston2010}. One consequence of such an 
optimisation of the 
variational free-energy is that it provides a bound on surprise and enables 
system's resilience while driving it to an equilibrium. 

Free energy principle works well when functional minimization works well. That 
is, when dealing with a single basin descent, as also the standard Bayesian 
inference procedure does. 
It is well known that this is true for all gradient descent methods which 
essentially describe approach towards equilibrium, the minimum. But, when 
one encounters multiple coexisting basins of attraction, stochastic terms have 
to enter into the picture. Then the need to extend the classic Bayesian 
framework arises \cite{BIB_Frontiers2022}. The situation is analogous 
of the metastable state-transitions in statistical thermodynamics driven by 
stochastic fluctuations, a typical far-from-equilibrium process. Indeed, this 
analogy has been noted  \cite{Friston2012,Friston2010} and, normative 
connections with formal analogies have been established with self-organization, 
autopoiesis, second order cybernetics and other theories with similar 
minimization challenges. 

\noindent {\bf Extending the Bayesian Approach:}
Parallel thoughts about the Bayesian inference in 
the presence of missing or incomplete data led statisticians and 
data-scientist to consider the inverse problem of the Bayesian theorem and 
supply the converse Bayesian theorem \cite{Ng2014,Tan2009}. 
In other words as the classic Bayes' theorem provides a better estimate, called 
`the posterior', for an original hypothesis expressed as a probability 
distribution, called `the prior', taking in consideration given data, called 
`the likelihood'; so the converse Bayes theorem provides a prior 
distribution that is compatible with the given likelihood furnished by the data 
(even if there are some of them missing: hence the name `missing data 
problems' \cite{Tan2009}) for a  given knowledge of the posterior. 

F. T. Arrechi, another pioneer of nonlinear science, made a further 
breakthrough when he explained neurophysiological data from ambiguous 
pictures, the Necker Cube in particular, based on an argument of Quantum 
Cognition theory and proposed a scheme for interlacing Bayesian 
and Inverse Bayesian (BIB) inference \cite{Arechi2011,Arechi2003}, in a closed 
feedback loop, see also his chapter contributed to \cite{JSN_GN_VB_2015}.
Further studies \cite{BasiosGunji2017,GunjiBasios2017,GunjiBasios2016} revealed 
that the underlying logic of Arrechi's experiments is a quantum-type of 
non-Boolean logic (an orthomodular lattice of propositions) in agreements with 
the tenets of the theory of Quantum Cognition \cite{QC2010,QC2012,Aerts}.

One can start by denoting $d$ and $h$ the variables representing data and 
hypotheses, respectively. The conditional probabilities
$P(h|d)= \frac{P(d,h)}{P(h)}$ and  $P(h|d)=P(d,h)P(d)$ give via Bayes Theorem:
$P(h|d)P(d)=P(d|h)P(h)$. Since for changing hypotheses from a set $H=\{h_k, 
k=1,2,\dots N\}$ at each time-step $t$ we have $P(d)=\sum_k P(d|h_k)P(h_k)$,
and we obtain the iteration process, indexed with the time-step $t$:
\begin{equation}
P^t(h|d)=P^t(d|h)P^t(h)\sum_k Pt(d|h_k)P^t(h_k) \Rightarrow\ 
P^{t+1}(h)=P^t(h|d)
\end{equation}
The data $D= \{d_k,k=1,2,\dots N\}$ and hypotheses $H=\{h_k, 
k=1,2,\dots N\}$, are analogous to external stimuli and their 
internal representations, or categories. This is the classical Bayes 
inference compatible with the free energy minimization principle.
Expressed in am operator form as:
\begin{equation}
P^{t+1}(h) = {\bf B } P^t(h|d)
\end{equation}
   \begin{figure}[h]
   \centering
   \includegraphics[width=7.0cm]{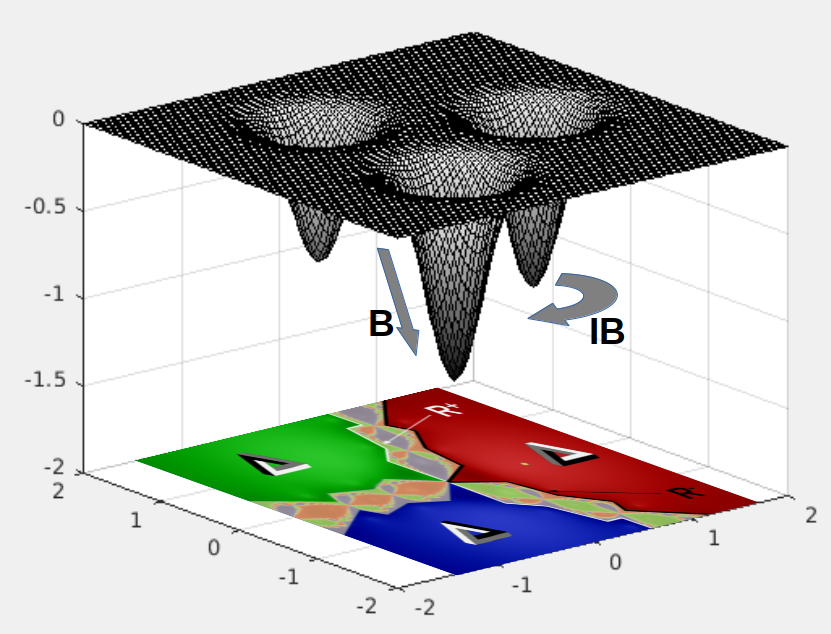}
   \caption{Bayesian Inference ({\bf B}, straight arrow) amounts to descending 
to a given basin, i.e. a category, or a hypothesis. Inverse 
Bayesian ({\bf IB}, curved arrow) inference amounts to hopping and 
switching among different basins/hypotheses.}
   \label{fig:BIBbasin}
   \end{figure}
The Inverse Bayesian inference (IB) can be expressed, also formally, as:
\begin{equation}
P^{t+1}(d|h) = {\bf IB } P^t(d) \mbox{, eq.(3) \& Bayes theorem, 
} \Rightarrow
P^{t+1}(h) = {\bf B }\diamond{\bf IB} P^t(h)
\end{equation}
\noindent dropping indexes for clarity and where the operator $\diamond$ 
denotes a special, non-deterministic, composition, respecting the tenets of the 
converse Bayesian theorem \cite{Ng2014,Tan2009}. One way t do this is be letting 
the joint probability between a hypothesis and data to be transformed into a 
binary relation $ R\subseteq H \times D$ such that $(h, d) \in R$ if $P^t(d, h) 
> \theta$; otherwise, $(h, d) \notin R$. Here $\theta$ is a threshold 
probability derived from the measure of the coarse-graining partition enclosed 
by the set $(R^+) - (R^-)$, e.g. as in the Fig.\ref{fig:Trillusion} and 
Fig.\ref{fig:BIBbasin}. Once a 
binary relation between a hypothesis and data is established, one can estimate 
a non-Boolean logical structure with respect to an, orthomodular, lattice.
In particular, a lower approximation on
hypotheses and an upper approximation on data, furnish a `Rough Set 
Approximation' \cite{GunjiBasios2016}. The iterative, non-algorithmic, part of 
the inverse Bayesian inference operator, {\bf IB}, is not simply `solving for 
P(h)' in Bayes’ formula of eqs.(2),(3). As with the Converse Bayes theorem, a 
correctly chosen functional space with proper convergence topology is crucial 
\cite{Ng2014}. 

This picture of a landscape of data and hypotheses with multiple coexisting 
attractors is a challenge for any deterministic minimization process and, 
hence, also for free energy principle schemes. It is reminiscent of 
out-of-equilibrium self-organization as it brings forth the role of stochastic 
fluctuations in state transitions and/or other non-deterministic factors.

\begin{figure}[h]
\centering
\includegraphics[width=9cm]{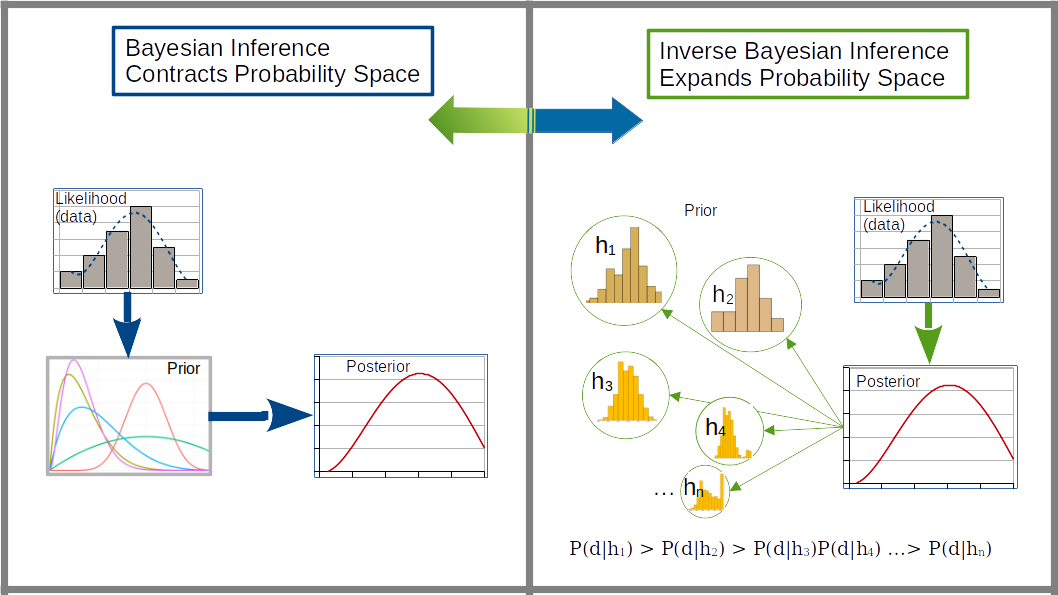}
\caption{The extended Bayesian (BIB) feedback loop. Left: its classical 
Bayesian inference branch. Right: the Inverse Bayesian inference branch 
with its extended probability distributions space that provide a set of 
competing hypotheses.}
\label{fig:BIB-new}
\end{figure}
Nevertheless, probabilistic strategies and 
‘on the fly’ construction of step-by-step solutions can work even if the 
process can be characterized as non-deterministic, or non-Turing-computable, 
in the normative sense of the words. Fig.\ref{fig:BIB-new}  
illustrates these two Bayesian Inference processes.

Again, here, we encounter the `exploit-explore' dual feedback loop as an 
instrumental and distinct feature of biological information 
processing. Now reflected in the probability space contraction/
expansion interplay during BIB inference:
\begin{tcolorbox}[width=12cm]
{\bf Bayesian Inference}: exploiting probability space\\
Given: Likelihood \& Prior distribution $\longmapsto$
Posterior distribution\\
{\bf Inverse Bayesian Inference}: exploring probability space\\
Given: Likelihood \& Posterior distribution 
$\longmapsto$ Prior distribution
\end{tcolorbox}


\section{Outlook \& Forthcoming Experiments}
\label{sec:conclusion}
Apart from elucidating the logical, non-Boolean, structure of apprehension and 
judgement and the correspondence with the conceptual framework of Quantum 
Cognition, so far BIB has been successfully been implemented in explaining the 
appearance of fractal-type Levy flight super-diffusion and other aspects of 
collective behaviour of swarms 
\cite{GunjiSwarm2018,Shinohara2020,GunjiSwarm2021,Shinohara_et_al_2022}. These 
are typical macroscopic biological processes. 

Yet, as new technologies emerge, we are now able to describe the dynamics of 
neural morphology with spatial resolution down to the nanoscale level. 
Nano-electromechanical vibrations became recently a powerful tool to 
investigate the role of oscillations, and noise trait, in the functioning of 
single neurons and the interaction with the environment \cite{Chialvo2010} (e.g. 
effects of drugs, motor activity etc.). 

In this context, experiments aiming at the analysis of such nano-vibrations of 
neurons have been designed and developed. We currently consider, the simplest 
complex system of neurons, consisting of just three neurons. This set-up allows 
the investigation of their collective behaviour in presence of different 
stimuli \cite{COMASAN}. The ultimate goal is to infer what processes make 
the simplest complex alive neuronal network to act as a collective entity.

Along side and in complementing classical deterministic modeling of 
synchronization modes of three coupled neurons \cite{Shilnikov2022}, BIB is 
expected to uncover other, not so commonly studied phenomena, in the simplest 
collective of neurons that nevertheless exhibit very complex behaviours, even 
infer new ones.
\\
\noindent{\bf Acknowledgments:}
GP-Y was supported by JSPS Topic-Setting Program to Advance Cutting-Edge 
Humanities \& Social Sciences, JPJS00120351748. P-FM \& VB acknowledge 
partial support from the Italian CNR Foresight Institute's program and from the 
Service de Physique des Systèmes Complexes et Mécanique Statistique of ULB.


\end{document}